\documentclass[conference]{IEEEtran}
\usepackage{cite}

\ifCLASSINFOpdf
   \usepackage{graphicx}
   \graphicspath{{img/}}
   \DeclareGraphicsExtensions{.pdf,.jpeg,.png}
\else
  \usepackage[dvips]{graphicx}
  \graphicspath{{../eps/}}
  \DeclareGraphicsExtensions{.eps}
\fi
\usepackage{amsmath}
\usepackage{array}
\usepackage{url}

\usepackage{kantlipsum}

\usepackage{pgfplots}
\usepgfplotslibrary{units}

\begin{document}
%
\title{A Framework for Application-aware Networking by Delegating Traffic Management of SDNs}

\author{\IEEEauthorblockN{Hamidreza Almasi}
\IEEEauthorblockA{Department of Computer Engineering\\
Sharif University of Technology\\
Tehran, Iran\\
Email: almasi@ce.sharif.edu}
\and
\IEEEauthorblockN{Hossein Ajorloo}
\IEEEauthorblockA{School of Computer Science\\
Institute for Research in Fundamental Sciences (IPM)\\
Tehran, Iran\\
Email: ajorloo@ipm.ir}}


%


\maketitle

\begin{abstract}
Presence of a logically centralized controller in software-defined networks enables smart and fine-grained management of network traffic. Generally, traffic management includes measurement, analysis and control of traffic in order to improve resource utilization. This is done by inspecting corresponding performance requirements using metrics such as packet delay, jitter, loss rate and bandwidth utilization from global network view. There has been many works regarding traffic management of software-defined networks and how it could help to efficiently allocate resources. However, the vast majority of these solutions are bounded to indirect information retrieved within the border of ingress and egress switches. This means that the three stage loop of measurement, analysis and control is performed on switches in between this border while the traffic flowing in network originates from applications on end hosts. In this work, we present a framework for incorporating network applications into the task of traffic management using the concept of software-defined networking. We demonstrate how this could help applications to receive desired level of quality of service by implementing a prototype of an API for flow bandwidth reservation using OpenFlow and OVSDB protocols.
\end{abstract}


%
\IEEEpeerreviewmaketitle

\section{Introduction}
The term SDN was first introduced in an article which was describing the OpenFlow project at Stanford University\cite{History}. OpenFlow stemmed from observing the fact that most new networking community ideas went untried and untested, so it was considered a means for researchers to let them test their experimental protocols in a production network \cite{OpenFlow}. It enabled programmability of network by setting rules in switches flow tables. Since then, the definition of SDN has expanded and OpenFlow is just one possible data plane or southbound API. Furthermore, SDN relies on control and data plane separation but does not reduce to just this, but in general, SDN is all about abstraction provided by forwarding devices with well-defined instruction sets, carefully designed network operating systems called controllers along with data and control plane separation by using appropriate southbound and northbound APIs \cite{Survey}.

The controller in software-defined networks maintains a global network view of traffic flows, link connectivity states, network resource allocation, etc. The control plane in these networks installs arbitrary forwarding and monitoring rules in TCAM entries of SDN switches to be matched on by data plane. This provides near-optimal, fine-grained decision making for traffic management tasks such as load balancing, fault tolerance, topology updates and traffic analysis compared to sub-optimal, coarse-grained decisions made in legacy networks.

Measuring and controlling QoS parameters like packet delay, jitter, loss rate and throughput is an inherent part of traffic management tasks and an important input to many decisions in computer networks. There has been several approaches towards maintaining these parameters in different types of networks. With more visibility into network resources, SDN introduced more efficient approaches \cite{TESurvey}. However, these solutions are limited to knowledge of measurements performed on flow table entries within ingress and egress switches area. In other words, knowledge of controller about network traffic is confined to statistics of the rules it eventually had to install in order to forward the traffic towards destination while the traffic originates from end host applications.

In this paper, we argue that in some environments such as data centers, these network applications better know the nature and anticipated QoS of the traffic they are generating and thus might be able to help the controller to treat the traffic in a more efficient manner. When network application programmers are sending data in some parts of their code development process, they could ask the network controller through an API for special QoS treatment tailored to the nature of the data and the controller could investigate the possibility of providing the service and respond accordingly. By using this framework, some portions of traffic management are delegated to applications and the network becomes more application-aware. This is not straightly possible in legacy networks due to distributed control. We will discuss that DiffServ \cite{DiffServRFC} does not provide fine-grained guarantees some applications might expect. However, the logically centralized controller in software-defined networks enables more direct monitoring and control on network resources while being a designated authority who the applications could talk to through an API.

The remainder of this paper is organized as follows: Section \ref{relatedWork} discusses related work on traffic management. Section \ref{Framework} presents the suggested framework and Section \ref{Prototype} continues by presenting a use case implementation for guaranteed minimum bandwidth. Section \ref{Evaluation} evaluates the prototype and Section \ref{Conclusion} concludes the paper.
%
%

\section{Related Work} \label{relatedWork}
There are several research works in the field of quality of service maintenance and using application layer information both in legacy and software-defined networks. We tend to discuss more practical solutions here.

DiffServ \cite{DiffServRFC} is a simple and scalable mechanism for classification and management of traffic to provide QoS in IP networks. This method can bring low-delay service for latency-sensitive traffic while using best-effort delivery for web and file transfer purposes. This class-based, coarse-grained mechanism performs all traffic classification at DiffServ domains borders and conceals the complexity from network core and unlike IntServ \cite{IntServRFC} and RSVP \cite{RSVPRFC}, it does not require reservation and end-to-end signaling, but on the other hand, how each router deals with DSCP bits depends on its configuration and therefore it is hard to predict end-to-end behavior.

Hedera \cite{Hedera2}, is a dynamic and scalable solution for flow scheduling avoiding ECMP limitations. It has a global view of traffic demands and routing information. Hedera scheduler collects flow information from switches, computes collision free paths and instructs switches to conduct traffic along those paths in a control loop. This periodic polling of traffic information or packet sampling leads to high monitoring overheads which incurs significant switch resources consumption and long event detection times. Mahout \cite{Mahout} suggests to use an end host shim layer which marks the packets of elephant flows. The switches are configured to forward marked packets to the Mahout controller. This resolves switch monitoring overhead issue. MicroTE \cite{MicroTE} notes that recent proposals for traffic engineering in ISPs operate at long time-scales and based on measurements \cite{BensonIMC}, data center traffic is naturally bursty and unpredictable at such long time-scales but a significant amount of traffic is predictable at short time-scales. It assigns servers for aggregating traffic statistics and sending summarized traffic matrix to the network controller which takes appropriate routing decisions for both predictable and non-predictable traffic. Yet, using either shim layer or top-of-rack aggregator may not deliver fine-grained control to individual network applications.

In \cite{AutoQoS}, a QoS controller is presented which can create network slices and provision them dynamically to satisfy performance requirements across applications it has assigned their traffic to those slices. Network administrators specify high level slice specifications and controller reserves network resources accordingly. Their main difference to our approach is that we delegate traffic management of our software-defined network to its applications and network administrators specify high level policies and limits for those authorized applications rather than slice specifications and we do this by developing a northbound API rather than extending southbound APIs such as OpenFlow since there is potentially more work to do in this part.

PANE \cite{PANE} uses the concept of share tree to perform similar tasks, but the lack of resource monitoring for maintaining a network snapshot limits its applicability in environments where not interested users coexist. Authors in \cite{PolicyRef} focus on SLA policy refinement applied to SDNs while this work focuses on delegating a part of policy enforcement process to applications.

\section{Proposed Framework} \label{Framework}
As mentioned, participating network applications in traffic management task of a software-defined network could better satisfy their service performance requirements. Here, we present the suggested framework which enables the controller to be aware of how the developer expected the network to behave with the traffic coming from the application.

\subsection{Overview}
\begin{figure}[!t]
\centering
\includegraphics[width=3.1in]{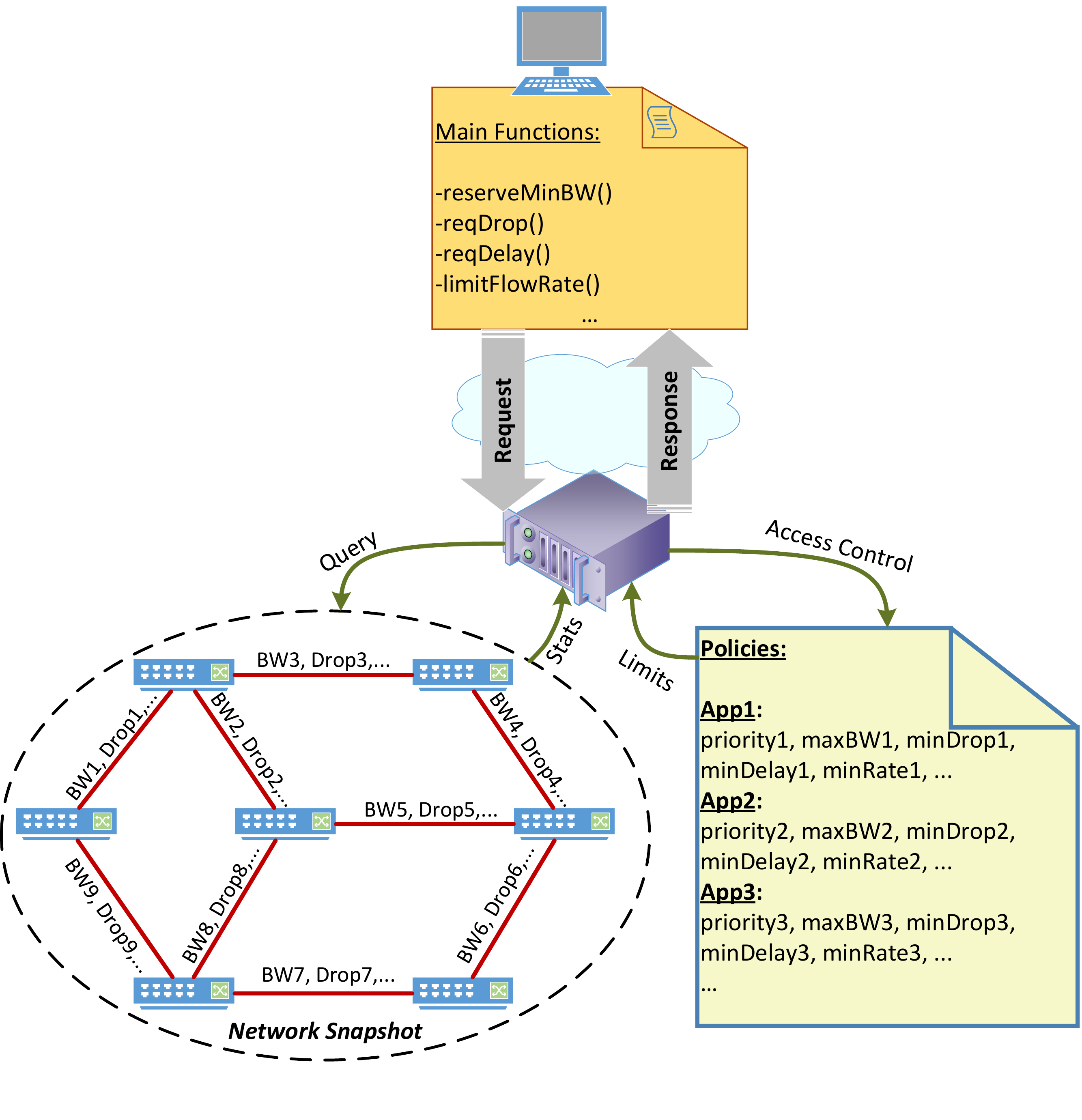}	
\caption{Framework overview}
\label{fig_overview}
\end{figure}
Using this framework, network application developers use an API to connect to destination with a differentiated socket when they require special treatment from the network. When a function is called from this API, there will be a negotiation phase with the controller which is constantly updating a snapshot of network resources status. The controller would then accept, reject or revise the request and then implements the behavior as rules in case it conforms to policies provided by network administrator and required resources are available. Finally, in case of request acceptance, the controller reserves resources for that flow for some time and then the connection is made as usual. Otherwise it enforces a revision or rejects the request and the connection is made without any special treatment. Requests from applications are put in a queue and processed according to application priority or on a first-come, first-served basis in case of equal priorities. An overview of this framework with its abstract interactions is depicted in Fig.~\ref{fig_overview}.

Four instances of API functions with their corresponding policies and network statistics data structures are shown in the figure.
\begin{table*}[!t]
\renewcommand{\arraystretch}{1.3}
\caption{API Specifications}
\label{table_specs}
\centering
\begin{tabular}{|c|c|m{0.31\linewidth}|m{0.25\linewidth}|}
\hline
\multicolumn{1}{|c|}{\bfseries Function} & \multicolumn{1}{c|}{\bfseries Specific Argument} & \multicolumn{1}{c|}{\bfseries Policy Check Expression} & \multicolumn{1}{c|}{\bfseries Description}\\
\hline
\texttt{reserveMinBW()} & \texttt{minExpectedBandwidth} &
${(minExpectedBandwidth \leq maxBW)} \land {(\exists p \in simplePaths(G, source, dest)}\mid {\min\limits_{\forall l \in L_p} (C_l-U_l) > minExpectedBandwidth})$
& Guarantees minimum bandwidth a flow receives. $L_p$ is the set of links forming path $p$. $C_l$ and $U_l$ are capacity and utilization of link $l$ respectively.\\
\hline
\texttt{reqDrop()} & \texttt{maxExpectedDropRate} &
${(maxExpectedDropRate \geq minDrop)} \land {(\exists p \in simplePaths(G, source, dest)}\mid {\max\limits_{\forall i \in I_p} (D_i) \leq maxExpectedDropRate})$
& Enforces an upper bound for the drop rate a flow might suffer. $I_p$ is the set of network interfaces along path $p$. $D_i$ is the drop rate of interface $i$.\\
\hline
\texttt{reqDelay()} & \texttt{maxExpectedDelay} &
${(maxExpectedDelay \geq minDelay)} \land {(\exists p \in simplePaths(G, source, dest)}\mid {Delay(p) \leq maxExpectedDelay)}$
& Enforces an upper bound for the delay a flow might encounter. Delay($p$) calculates average latency the flow packets encounter along path $p$.\\
\hline
\texttt{limitFlowRate()} & \texttt{maxFlowRate} &
${(maxFlowRate \geq minRate)} \land {(\exists p \in simplePaths(G, source, dest)}\mid {U_l \underset{\forall l \in L_p}< C_l})$
& Enforces traffic contract between tenants of flow source and destination. $C_l$ and $U_l$ are capacity and utilization of link $l$ in path p respectively.\\
\hline
\end{tabular}
\end{table*}
Their common arguments are \texttt{socketType}, \texttt{destIP} and \texttt{destPort}. Their specific argument and policy check expression is indicated in Table \ref{table_specs}. Method simplePaths(G, source, dest) operates over graph $G=(V, E)$ from a topology discovery module feeding network snapshot where $V$ is the set of OpenFlow switches and $E$ is the set of links interconnecting them and returns acyclic paths from source to destination. Note that these functions are only some samples of how QoS enforcement tasks could be plugged into this framework and the generality of framework allows for other forms of functions and parameters.
We next describe how the controller becomes aware of application intents for traffic.

\subsection{Application-Controller Interaction}
\begin{figure*}[!t]
\centering
\includegraphics[scale=0.6]{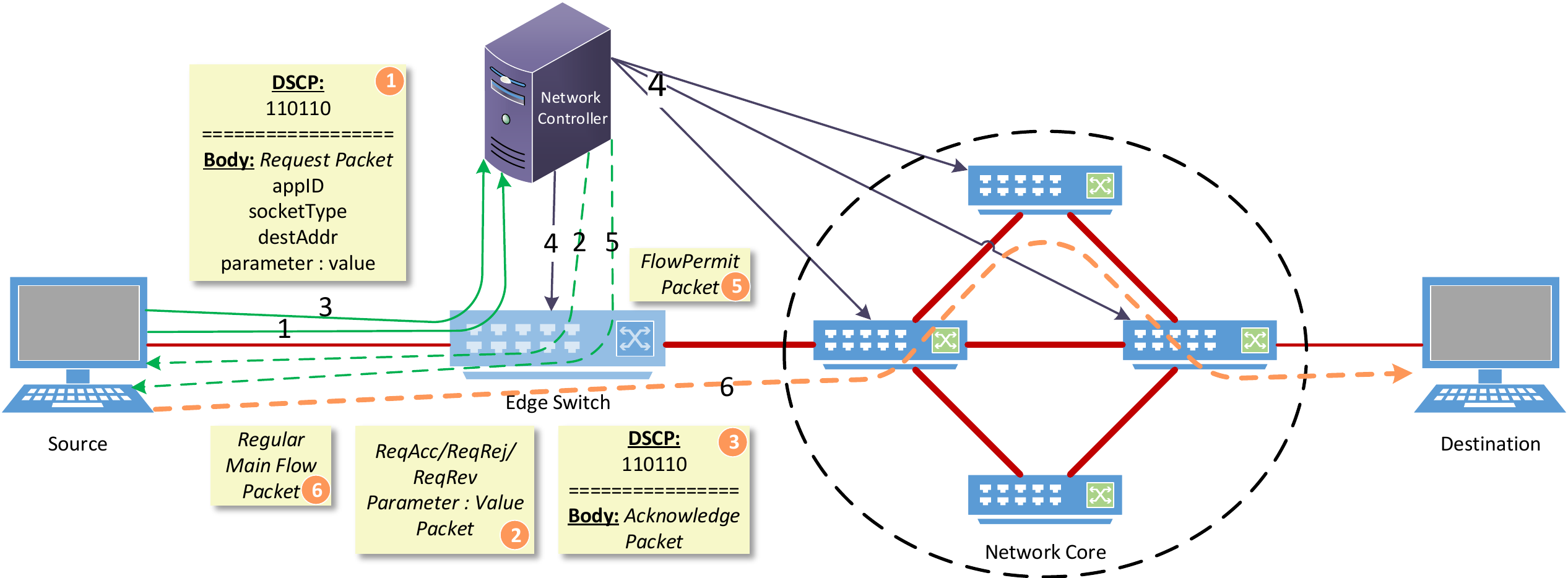}
\caption{Steps of application-controller interaction until traffic flow}
\label{fig_interaction}
\end{figure*}
When an application decides to send some traffic under special QoS requirements, it has to notify the controller about those requirements and also receive the response to know whether some special treatment is granted or not. Fig.~\ref{fig_interaction} illustrates steps of this interaction. For this purpose, in API function calls, before a socket to the destination is created, there is a negotiation phase with the controller in which a request packet containing application and destination identifiers, socket type and intended QoS requirements in the body is prepared. Here is what happens next:
\begin{itemize}
    \item A TCP socket to the controller is created with packet DSCP bits set to \texttt{110110}, a value not in commonly used values in \cite{DiffServRFC} for consistency. We also pre-install a rule after discovery of each switch to match on these specific DSCP bits destined to the controller and forward those packets for further processing (step 1).
    \item When the controller module responsible for this API receives this type of packet, it prepares a response according to what we described earlier, installs a rule in switches on path to the requesting application and forwards the response (step 2).
    \item The requesting application finally acknowledges the response or revises the request based on specified limits announced by the controller and acknowledges revision request (step 3).
    \item The controller installs necessary rules in switches along the path with feasible resources (step 4).
    \item Based on rules set in step 2, the application-side API module which is waiting for resource allocation completion signal from the controller is notified about this (step 5).
    \item Finally the application-side module creates its socket to the destination as before and main data flow starts by the application (step 6).
 \end{itemize}
\subsection{Monitoring Network Resources}
Generally, there are two approaches to network monitoring and measurement. In active monitoring, some measurement traffic is injected into the network. For example a probe sent to the network will trigger a response and one could analyze the response to infer properties of the network. In passive monitoring, the traffic that naturally exists in the network is observed.

For OpenFlow networks \cite{OFSwitchSpec}, active monitoring translates to relying on \texttt{OFPT\_MULTIPART\_REQUEST} and \texttt{OFPT\_MULTIPART\_REPLY} messages exchanged between switches and the controller. While polling switches for statistics at arbitrarily high rates may provide accurate view of the network, this could incur significant overhead and impose scalability problems in large, heavily loaded networks. On the other hand, passive monitoring in OpenFlow networks means listening to asynchronous messages \texttt{OFPT\_PACKET\_IN}, \texttt{OFPT\_FLOW\_REMOVED} and \texttt{OFPT\_PORT\_STATUS} to infer some properties like link utilization and topology updates. Although passive method has less overhead, but it sacrifices accuracy and also not all QoS metrics can be inferred in this manner.

The monitoring methods plugged into this framework should provide a real-time, accurate snapshot of network so that the framework could properly decide how to respond to applications requests. We utilize adaptive, active statistics collection method similar to the algorithm in \cite{Payless} which trades between accuracy and overhead and apply it to ports along simple network paths from source to destination. It adapts polling frequency to activity of the ports or amount of traffic that flows contribute to them and calculates link utilization for bandwidth guarantees and traffic shaping and also number of dropped packets and errors occurred. Regarding path delays, it is not feasible to apriori maintain all-pairs delays. This is because total delay is composed of several components such as processing, queueing and transmission delays which are added up together and cannot be independently captured and saved before a probe destined for that specific path is sent by \texttt{OFPT\_PACKET\_OUT} and \texttt{OFPT\_FLOW\_MOD} messages and also due to possible large number of simple paths between source and destination and limited size of flow tables, such a solution would not be practical. In this case we suggest to set the rules for navigating the probe towards the egress switch and send it upon the receipt of this kind of request. 
\section{Prototype Implementation} \label{Prototype}
We implemented bandwidth reservation and traffic shaping scenarios as instances of our proposed framework usage over Ryu which is a widely used component-based SDN controller \cite{RyuWeb}. Ryu is a modular controller. Each module registers itself to listen for specific events (i.e., messages from other modules) and upon receipt of events, puts them in a FIFO queue. It also has a thread for event processing which keeps dequeuing from this receive queue and calling appropriate event handler. We have a module that tracks topology changes based on LLDP packets plus a monitoring module that maintains links utilization and ports status. Another module processes application requests for validity based on network snapshot information and administration policies and evaluates whether the request complies with its criteria in Table \ref{table_specs}. For example, regarding minimum bandwidth reservation scenario it checks if the requested amount is less than the maximum bandwidth considered for that application and there is a path which can accommodate the flow. Finally a module implements the behavior in data plane.

We used OpenFlow version 1.3 \cite{OFSwitchSpec} for communication between the control plane and Open vSwitch \cite{OVSWeb} data plane in Mininet \cite{MininetWeb} test and prototyping environment. Mininet uses lightweight virtualization meaning a host is a shell process moved into its own network namespace. Mininet hosts have their own virtual Ethernet interface(s) \cite{Mininet}. For generating test traffic we used iPerf \cite{iPerfWeb} and for graph algorithms we used NetworkX package \cite{NetworkXWeb}. All scripts were written in Python.

To implement bandwidth reservation and traffic shaping scenarios we used OVSDB \cite{OVSDBRFC} management interface as a complementary protocol to OpenFlow for configuration of switches, ports, queues and attachment of QoS policies to queues. We chose hierarchical token bucket queuing discipline and set configurations in JSON format.

\section{Evaluation} \label{Evaluation}
\begin{figure}[!t]
\centering
\includegraphics[width=3in]{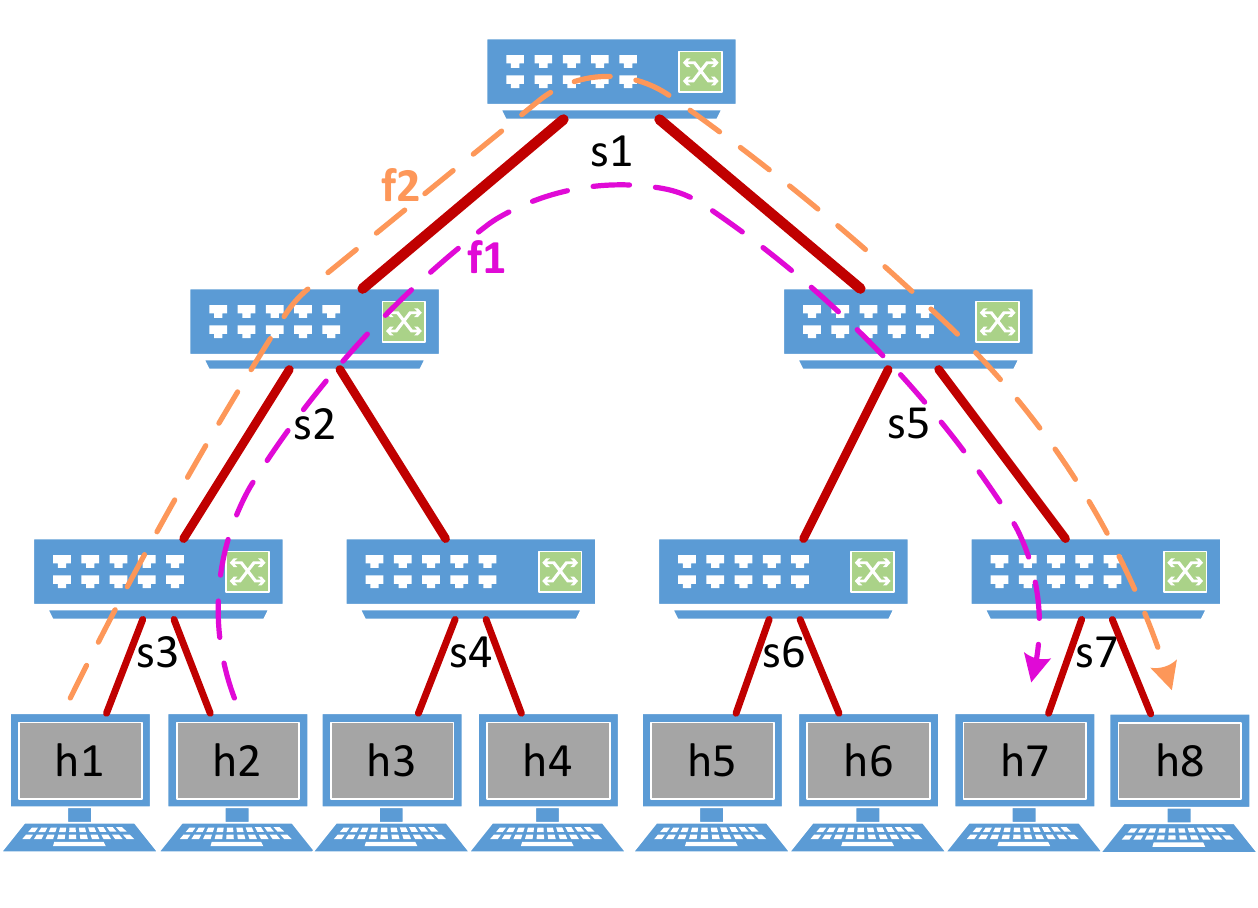}
\caption{Topology used for evaluating effectiveness of bandwidth reservation and traffic shaping scenarios. All links have 20Mbps capacity and intended minimum and maximum rate for flow f2 is 10Mbps.}
\label{fig_testTopo}
\end{figure}

We simultaneously evaluated the framework operation under the two implemented scenarios by requesting same upper and lower bandwidth limit through our API functions. A tree topology with 7 switches (\texttt{s1-s7}) and 8 hosts (\texttt{h1-h8}) was created with Mininet. All link capacities were set to 20Mbps. Two UDP flows \texttt{f1} originating from \texttt{h2}, destined to \texttt{h7} and \texttt{f2} originating from \texttt{h1}, destined to \texttt{h8} were created with iPerf. Flow \texttt{f1} was configured as a link throttling flow while \texttt{f2} was intended to receive 10Mbps bandwidth. Test environment is depicted in Fig.~\ref{fig_testTopo}.

The bandwidth perceived by \texttt{f2} was measured in two circumstances, with and without calling API functions, to study framework operation and API effectiveness. We configured iPerf to report the bandwidth for each flow in 0.5 second intervals for a 10 second time period. For evaluation purpose, upper and lower bounds bandwidth policies for the application generating \texttt{f2} were set in compliance with its request. As shown in Fig.~\ref{f2rate}, the average \texttt{f2} rate is 9.72Mbps with minimum variance when using our framework compared to the average of 5.2Mbps without using the framework. When the connection is made in traditional way, \texttt{f1} overwhelms \texttt{f2} by competing over shared links, but calling function \texttt{reserveMinBW()} before flow \texttt{f2} starts, makes the controller set some queues along the path to the destination.

For a graph $G=(V, E)$, a simple path is found in $O(V+E)$ using a modified DFS algorithm but since the number of simple paths might be as high as $O(n!)$ for a complete graph of degree $n$, we exploit a deepening approach to check shorter paths first. Furthermore, a multi-commodity flow problem arises if requests are not handled readily. Assuming that link $(u, v)$ has capacity $c(u, v)$, there are $k$ commodities $K_1, K_2, ...,K_k$ defined by $K_i = (s_i, t_i, d_i)$ where $s_i$ and $t_i$ are source and sick of commodity $i$ and $d_i$ is the bandwidth demand. The flow of commodity $i$ along link $(u, v)$ is $f_i(u, v)$ and the problem is finding an assignment of flows which satisfies the following constraints:
\begin{IEEEeqnarray}{c}
\sum_{i=1}^{k} f_i(u, v) \leq c(u, v)\\
\sum_{w \in V} f_i(u, w) = 0 \text{ when } u \neq s_i,t_i\\
\forall v, u: f_i(v, u) = -f_i(u, v)\\
\sum_{w \in V} f_i(s_i, w) = \sum_{w \in V} f_i(w, t_i) = d_i
\end{IEEEeqnarray}
which is NP-Complete while we are facing a time-critical decision. We thus confine to a first-fit heuristic for flow placement. Table \ref{table_time} indicates path computation time for flows generated in a $k$-ary fat-tree network topology introduced in \cite{Al-Fares}. In this topology which is an instance of Clos topology, there are $k$ pods, each containing two layers of $k/2$ switches with $k$ ports. So there are $k^2/2$ edge switches that we consider triggering points of \texttt{OFPT\_PACKET\_IN} messages. There are $(k/2)^2$ paths between any two hosts on different pods and for evaluation we assume inter-pod traffic, therefore our first-fit heuristic finds paths for bandwidth reservation scenario in $O(k^3)$. We simulate a performance test for the controller by sending a \texttt{OFPT\_PACKET\_IN} from a switch to the controller and waiting for a matching \texttt{OFPT\_FLOW\_MOD} and counting how many times this happens in a second. In each round, we perform this for each one of $k^2/2$ edge switches. We do the test for 15 rounds and average over total number of responses received and path computation times. As evident in Table \ref{table_time}, when the number of pods  (same as the number of switch ports) increases, the number of switches increases even more and this leads to a larger graph to search and more path computation time. Meanwhile, since a greater portion of controller's time is dedicated to graph computations, each event handling would take more time and less new flows could be serviced in each time period. But even for a network with $k=40$ which supports $k^3/4=16000$ hosts, this time is less than 2 milliseconds.
\begin{figure}[!t]
\begin{tikzpicture}
	\begin{axis}[use units,
    		x unit=s,
    		y unit=b/s,y unit prefix=M,
		xlabel=Time after start,
		ylabel=Rate,
		height=2.5in,
		width=3in,
		grid=major,
		legend style={cells={align=left}, at={(0.44,0.72)},anchor=west,font=\scriptsize}
	]

	\addplot coordinates {
		(0.25,9.74)
		(0.75,9.71)
		(1.25,9.74)
		(1.75,9.71)
		(2.25,9.71)
		(2.75,9.74)
		(3.25,9.71)
		(3.75,9.71)
		(4.25,9.74)
		(4.75,9.71)
		(5.25,9.71)
		(5.75,9.74)
		(6.25,9.71)
		(6.75,9.74)
		(7.25,9.71)
		(7.75,9.71)
		(8.25,9.74)
		(8.75,9.71)
		(9.25,9.71)
		(9.75,9.74)
	};
	\addlegendentry{Using framework API}
	\addplot coordinates {
		(0.25,9.64)
		(0.75,9.55)
		(1.25,9.36)
		(1.75,6.30)
		(2.25,6.35)
		(2.75,4.87)
		(3.25,5.10)
		(3.75,4.73)
		(4.25,4.77)
		(4.75,4.87)
		(5.25,4.87)
		(5.75,4.87)
		(6.25,4.87)
		(6.75,4.87)
		(7.25,4.87)
		(7.75,4.87)
		(8.25,4.87)
		(8.75,4.87)
		(9.25,4.73)
		(9.75,4.68)
	};
	\addlegendentry{With no use of\\ framework API}

	\end{axis}
\end{tikzpicture}
\caption{Bandwidth perceived by UDP flow f2}
\label{f2rate}
\end{figure}
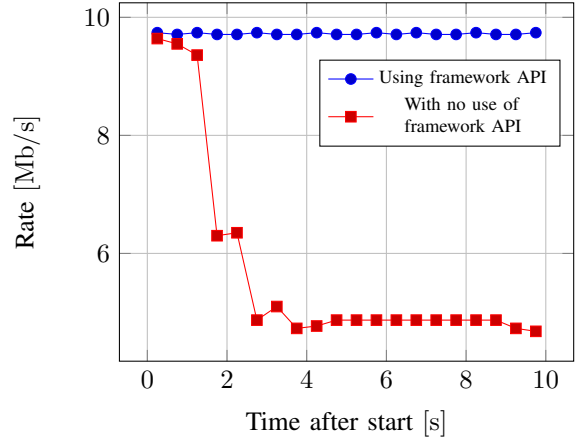
\begin{table}[!t]
\renewcommand{\arraystretch}{1.3}
\caption{Controller Performance and Processing Time for Bandwidth Reservation Scenario under Different Network Sizes}
\label{table_time}
\begin{tabular}{|c|c|c|c|}
\hline
\multicolumn{1}{|m{0.15\columnwidth}|}{\centering {\bfseries Number of Pods $(k)$}} & \multicolumn{1}{m{0.15\columnwidth}|}{\centering {\bfseries Number of Edge Switches $(k^2/2)$}} & \multicolumn{1}{m{0.25\columnwidth}|}{\centering {\bfseries Average Path Computation Time $[s]$}} & \multicolumn{1}{m{0.2\columnwidth}|}{\centering {\bfseries Average Number of New Flows Handled $[1/s]$}}\\
\hline
$4$ & 8 & $4.63466085052\mathrm{e}{-6}$ & $2769.29$\\
$8$ & 32 & $2.29313795225\mathrm{e}{-5}$ & $2726.00$\\
$12$ & 72 & $5.93899789978\mathrm{e}{-5}$ & $2588.71$\\
$16$ & 128 & $0.000132252089811$ & $2374.65$\\
$20$ & 200 & $0.000241951589853$ & $2256.31$\\
$24$ & 288 & $0.000388239264119$ & $1967.76$\\
$28$ & 392 & $0.000587952680526$ & $1821.77$\\
$32$ & 512 & $0.000849910279676$ & $1729.95$\\
$36$ & 648 & $0.00125786274914$ & $1302.46$\\
$40$ & 800 & $0.00170276373988$ & $1135.03$\\
$44$ & 968 & $0.00233802724793$ & $686.76$\\
\hline
\end{tabular}
\end{table}
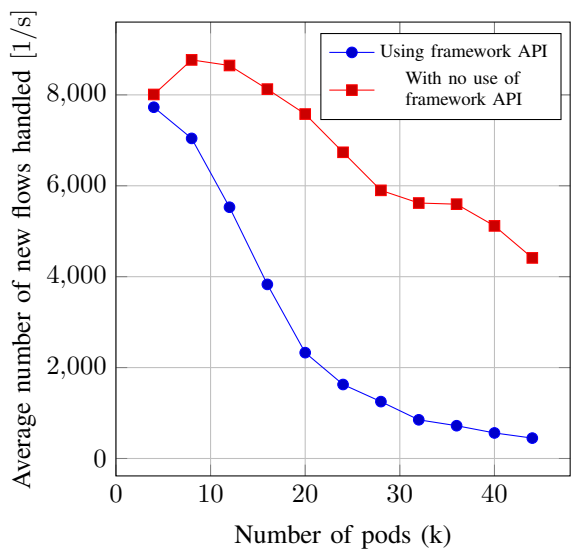
\begin{figure}[!t]
\centering
\begin{tikzpicture}
	\begin{axis}[use units,
    		y unit=1/s,
		xlabel=Number of pods (k),
		ylabel=Average number of new flows handled,
		height=3in,
		width=3in,
		grid=major,
		legend style={cells={align=left},font=\scriptsize,at={(0.45,0.88)},anchor=west}
	]

	\addplot coordinates {
		(4,7728.19)
		(8,7042.52)
		(12,5528.31)
		(16,3830.26)
		(20,2328.31)
		(24,1627.77)
		(28,1251.24)
		(32,851.05)
		(36,721.10)
		(40,562.06)
		(44,448.53)
	};
	\addlegendentry{Using framework API}
	\addplot coordinates {
		(4,8009.05)
		(8,8769.51)
		(12,8646.43)
		(16,8126.54)
		(20,7577.70)
		(24,6736.77)
		(28,5899.62)
		(32,5620.55)
		(36,5595.16)
		(40,5117.59)
		(44,4413.05)
	};
	\addlegendentry{With no use of\\ framework API}

	\end{axis}
\end{tikzpicture}
\caption{Controller's ability to handle new flows under different network sizes. Worst-case condition is assumed for API usage.}
\label{overhead}
\end{figure}

In order to get a better sense of the overhead our graph computations impose on the controller, we also compared the average number of new flows handled by the controller in two situations, when the controller is barely running a packet\_in handler to respond with a simple \texttt{OFPT\_FLOW\_MOD} with no hesitation and when an algorithm of $O(k^3)$ is run by the handler before responding. Fig.~\ref{overhead} demonstrates the results for different network sizes. As depicted in this figure, the controller remains performant even in highest traffic demands. Note that we have considered the worst-case scenario for overhead analysis because the number of times a graph algorithm is required to run, depends on the number of accepted requests from applications and we know that normally, in a MapReduce~\cite{MapReduce} shuffle for example, there is a processing time between requests. So, not all packet\_in events require graph computations while this is the case with our performance tests. The same reasoning applies to graph search operation where many graph searches terminate sooner than we assumed. Anyway, as mentioned earlier, there is room for improvement of monitoring and controlling algorithms used in this framework when specific knowledge of traffic and communication patterns is present.

\section{Conclusion} \label{Conclusion}
This paper describes a framework for delegating traffic management tasks such as QoS enforcement to end-host applications in a software-defined environment. This is a move towards more practical application-aware networking enabled by centralized control in SDNs. With this framework, network administrators could specify policies for authorized applications and network application developers could use an API to inform the controller of their specific requirement for a flow. Several monitoring and control mechanisms can be incorporated into the framework. Evaluation of our implemented prototype for minimum bandwidth reservation and traffic shaping functions shows effective operation of this framework allowing integration of further traffic management tasks and network monitoring approaches.


%
%



\bibliographystyle{IEEEtran}
\bibliography{IEEEabrv,resources}
%
%
%

\end{document}